\documentclass[12pt]{iopart}
 \usepackage{graphics}
 \usepackage{epsfig}
 \usepackage[english]{babel}

\begin{document}

\title[Creation of the inverse population in the $^{229}$Th ground-state doublet]{Creation of the inverse population in the $^{229}$Th ground-state doublet by means of narrowband laser}

\author{E V Tkalya$^{1,2}$ and L P Yatsenko$^3$}

\address{$^1$ Skobeltsyn Institute of Nuclear Physics, Lomonosov Moscow State University, Leninskie gory, GSP-1, Moscow 119991, Russia}
\address{$^2$ Nuclear Safety Institute of Russian Academy of Science, Bol'shaya Tulskaya - 52, Moscow 115191, Russia}
\address{$^3$ Institute of Physics, National Academy of Science of Ukraine, Nauky av. 46, Kyiv 680028, Ukraine}
\ead{tkalya@srd.sinp.msu.ru}

\begin{abstract}
A new method for obtaining of inverse population between quadrupole sublevels of the ground state $5/2^+(0.0)$ and the isomeric state $3/2^+(7.6$ eV) of the $^{229}$Th nucleus in a dielectric crystal with a large band gap by means of external source of the VUV radiation is proposed. The method is based on an efficient depopulation of the upper sublevels of the ground state of $^{229}$Th by resonant photons from narrowband laser or broader tunable free electron laser. Sublevels of the isomeric state play the role of intermediate states. In addition, we have considered a case of excitation of the isomeric state (without creation of inverse population) by broadband source of light with the anti-Stokes scattering. The proposed scheme of optical pumping results in (1) inverse population of nuclear sublevels without using extremely low temperatures, and (2) shows a new way for the creation of the gamma-ray laser of optical range at the nuclear transition in $^{229}$Th (a) in crystals with isovalent substitution of host ions (eg Si$^{4+}$ ions in the SiO$_2$ crystal replaced by the $^{229}$Th$^{4+}$ ions) and (b) in crystals such as Na$_2$ThF$_6$, where the substitution is not necessary.
\end{abstract}


\submitto{Laser Physics Letters}
\maketitle

\par
\section{Introduction}
\label{sec:I}
\par

Laser on a nuclear transition is one of the most famous unsolved problems in modern physics \cite{Baldwin-97,Rivlin-07}. Recently it has been shown \cite{Tkalya-11}, that a breakthrough in the field of gamma-ray laser is possible using the doublet levels in the $^{229}$Th nucleus for working transition. This doublet consists of the ground state $J^{\pi}_{gr}(E_{gr})=5/2^+(0.0)$ and the isomeric state $J^{\pi}_{is}(E_{is})=3/2^+(7.6\pm 0.5$ eV) \cite{Beck-07}. Laser on the $^{229}$Th nucleus will have all features of the gamma-ray laser, but it will work at the wavelength of approximately $160$ nm, i.e. in the optical (VUV) region of the spectrum.

Creating such a device will allow us not only to prove the principal possibility of the amplification of electromagnetic radiation by the nuclear medium, but also make progress in solving many fundamental problems. There are some of them. 1. The effective pumping of the 7.6 eV nuclear transition in $^{229}$Th will create conditions: a) to check of the exponentiality of the decay law of an isolated metastable state at a long time (longer than 50$T_{1/2}$) \cite{Dykhne-98}, and b) to get the nuclear superfluorescence in the ensemble of the $^{229}$Th$^m$($3/2^+$, 7.6 eV) isomers (the collective emission of fluorescence by an ensemble of excited nuclei first considered in \cite{Terhune-65} is an analog of the cooperative spontaneous emission Dicke \cite{Dicke-54} for excited atoms or ions). 2. According to calculations of \cite{Flambaum-06} the effect of temporal variation of the fine structure constant $e^2$ and the dimensionless strong interaction parameter $m_q/\Lambda_{QCD}$ may be enhanced by 5--6 orders of magnitude in the 7.6 eV nuclear transition in $^{229}$Th. 3. The nuclear metrological standard of frequency \cite{Tkalya-96} is one more exciting application of the $^{229}$Th$^m$ isomer. It is assumed that the ``nuclear clock'' \cite{Peik-03} based on the low energy transition in $^{229}$Th will have unprecedented accuracy at the level of $1\times10^{-19}$ \cite{Campbell-12}. 4. In general, there is a unique situation for low-energy nuclear spectroscopy when nucleui in a host dielectric crystal interact directly with photons \cite{Tkalya-00-JETPL} (due to the large band gap of the host dielectric one can neglect the internal conversion and electronic bridge \cite{Tkalya-00-PRC,Tkalya-03}). Regarding the field of quantum optics and quantum computing, the isomeric state $3/2^+$(7.6 eV) in the $^{229}$Th nucleus, that could be directly interrogated by laser light, could served as a qubit with extraordinary features \cite{Raeder-11}. 5. Among other intersting applications of the unique doublet in the $^{229}$Th nucleus one can note the dependence of the $3/2^+$(7.6 eV) isomeric level half life (or the probability of the spontaneous decay) on the refractive index of the host dielectric \cite{Tkalya-00-JETPL,Tkalya-00-PRC}, and the M\"{o}ssbauer effect in the optical range \cite{Tkalya-11}.

The easiest way to populate the isomeric state $3/2^+$ with the energy of 7.6 eV in the $^{229}$Th nucleus is to use laser radiation tuned in resonance with the nuclear transition $5/2^+(0.0) \rightarrow 3/2^+(7.6$ eV) \cite{Tkalya-11}. However, it is not enough for achieving the inverse population. In the asymptotic limit (for long times of irradiation), the ratio of densities of the nuclei in the ground $n^{gr}$ and the isomeric $n^{is}$ states satisfies the condition $n^{gr}/n^{is} = \Lambda/\sigma\varphi + g$ \cite{Tkalya-11}. Here $\Lambda$ is the decay constant, which is connected with the half-life $T_{1/2}^{is}$ of unpolarized isomers $^{229}$Th$^m$($3/2^+,7.6$ eV) in the radiative transition $|3/2^+\rangle \rightarrow|5/2^+\rangle$ by the standard formula $\Lambda = \ln(2)/T_{1/2}^{is}$ (here we use the system of units $\hbar=c=1$), $\varphi$ is the flux density of the laser photons, $\sigma$ is the excitation cross section of unpolarized nuclei by unpolarized photons in the nuclear transition $|5/2^+\rangle \rightarrow|3/2^+\rangle$, and $g=(2J_{gr}+1)/(2J_{is}+1)$ is the statistical factor. Since for the considered doublet $g>1$, in the scheme of direct pumping the inverse population cannot be achieved for any value of $\sigma\varphi/\Lambda$.

A solution of this inverse population problem was proposed in \cite{Tkalya-11,Tkalya-12}. It is based on the following consideration. The ground and the isomeric states of the $^{229}$Th nucleus have magnetic and quadrupole moments. Therefore in nonzero magnetic and non-uniform electric field the degeneracy of these states is removed. For example, consider the interaction of the $^{229}$Th nucleus with internal crystal electric field. The non-uniform electric field leads to non-zero electric field gradient (EFG) at nuclear positions, which causes interaction with the nuclear quadrupole moments and thereby enables the splitting of the $5/2^+(0.0)$ and $3/2^+(7.6$ eV) nuclear levels. It is known that the splitting of nuclear states of $^{229}$Th in the $^{229}$Th:LiCaAlF$_6$ crystal \cite{Rellergert-10} is relatively large and can reach $10^{-5}$ eV \cite{Tkalya-11}. Provided that the condition for the nuclear spin relaxation (and the polarization of nuclei in magnetic field or the nuclear quadrupolar alignment in electric field) is fulfilled, the Boltzmann distribution in the population of the nuclear sublevels appears at an easily achievable temperature about 0.1 K \cite{Tkalya-11}. Inverse population then arises between the lower sublevels of the isomeric state and the upper sublevels of the ground state.

A serious problem of this method is the relaxation rate. In ionic crystals, including LiCaAlF$_6$, the relaxation of the nuclear spins and the establishment of the Boltzmann distribution of the population of nuclear sublevels will be extremely slow, because (1) there are no free electrons of such crystals, and (2) there are no optical phonons at cryogenic temperatures. The process of redistribution of the population of nuclear sublevels in ionic crystal can be accelerated, if one covers the dielectric sample by metal \cite{Tkalya-11,Tkalya-12}. In this case the relaxation of the nuclear spins will occur via the conduction electrons of the metallic covering. Unfortunately, the expected time of the establishment of the Boltzmann distribution of the populations of the nuclear sublevels (and, accordingly, the population of the lower sublevels of the isomeric state and the emptying of the upper sublevels of the ground state) is quite large and equals to tens or hundreds of days according to our preliminary estimates \cite{Tkalya-11,Tkalya-12}.

In this paper we propose another solution to the problem which is based on the technique of optical pumping (see, for example, the review \cite{Happer-72}) well known in atomic physics. It includes (a) emptying one of the upper sublevels of the ground state $5/2^+(0.0)$ via the sublevels of the isomeric state $3/2^+(7.6$ eV) by the radiation of narrow-band laser for a time of few half-lives of $T_{1/2}^{is}$, and (b) relatively fast (for a time of (0.1-0.5)$T_{1/2}^{is}$) excitation of the sublevels of the isomeric state by the radiation of narrow-band laser. The inverse population arises at the second stage between sublevels of the isomeric state and the ground state sublevel, which was exhausted at the first step. This method of producing of inverse population can be more convenient for experimental implementation.

Moreover, the proposed method of optical pumping can be crucially important for a solution of another problem. In crystals of the $^{229}$Th:LiCaAlF$_6$ type where the host ions (in this case, Ca$^{2+}$) are replaced by Th$^{4+}$ ions (so called non-isovalent substitute) there should be a compensation of additional charge (see in Sec.~\ref{sec:Depopulation}). Compensating ions (in the case of $^{229}$Th:LiCaAlF$_6$ these are two additional F$^-$ ions located near the Th$^{4+}$ ion) are positioned in the interstices of the crystal lattice, which can lead to the additional quasi inhomogeneous broadening of the $^{229}$Th emission line. The broadening can be avoided by choosing crystals with the isovalent host-guest substitution. For example, in $^{229}$Th:SiO$_2$ Th$^{4+}$ ions replace Si$^{4+}$ ions, or in Na$_2$$^{229}$ThF$_6$  $^{229}$Th$^{4+}$ ions partially substitute $^{232}$Th$^{4+}$ ions. The electric field gradient at Th$^{4+}$ sites depends on the crystal choice and can change substantially. If EFG is smaller than in $^{229}$Th:LiCaAlF$_6$, then as a consequence, the quadrupole splitting of the nuclear levels will be smaller too. For the relaxation mechanism from the work \cite{Tkalya-11} it represents a significant disadvantage because requires for the spin relaxation process one order lower temperature than in crystals with non isovalent substitution.

As for the scheme with optical pumping, it is obvious that it does not require extremely low temperatures of 0.01-0.1 K, which represents its great advantage. It shows a new way for the creation of gamma-ray laser of the optical range at the $^{229}$Th 7.6 eV nuclear transition in the Th:SiO$_2$ and NaThF$_6$ crystals and also increases the number of crystals which can serve as candidates for this goal.

VUV lasers, which we propose to use for optical pumping, is the rapidly developing field of current research \cite{Drake-06}. In the literature for the energy of the isomeric level we find the following approximate range: $E_{is}\approx 7.1$--$8.1$ eV \cite{Beck-07}. Therefore at the moment we can not tell which VUV laser will be suitable for the pumping of the $^{229m}$Th isomers. It is conceivable that we can use one of the available lasers. Their parameters are described in detail in the review \cite{Drake-06}. If not, it will be necessary to use a free electron laser, which has a broader tunability. In addition, one can use a powerful UV lamp or some other broadband UV source with a high spectral brightness in the range 150--170 nm, whose spectrum is processed with a monochromator. Thus later in this paper for brevity we call the two sources of radiation ``narrowband laser'' and ``broadband laser'' meaning different sources of UV radiation, which are suitable for our purposes.

\par
\section{Quadrupole splitting of the nuclear states}
\label{sec:Depopulation}
\par

The most promising scheme for the laser at the $M1$ $3/2^+(7.6$ eV)$\rightarrow5/2^+(0.0)$ nuclear transition in $^{229}$Th is the one which utilizes quadrupole splitting \cite{Tkalya-11} with a large electric field gradient at the $^{229}$Th crystal sites. In the following we will not consider the other scheme working on Zeeman sublevels, because it requires a very strong  external magnetic field and, consequently, is more difficult to implement.

\begin{figure}[]
\begin{center}
\epsfig{file=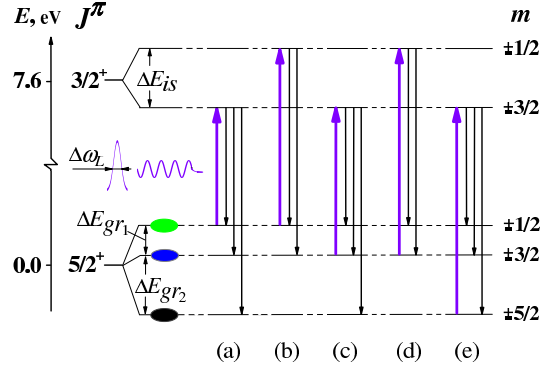,width=7.0cm}
\end{center}
\caption{Quadrupole splitting of the ground and the isomeric states of the $^{229}$Th nucleus in a crystal with a large electric field gradient and the initial population of the sublevels.  Deexcitation of the sublevels by narrowband laser: (a) and (b) depopulation of the $|5/2^+,\pm1/2\rangle$ sublevel by resonant laser radiation with $\Delta\omega_L\ll\Delta{E}_{gr_1}$ via the isomeric state sublevels $|3/2^+,\pm1/2\rangle$ and $|3/2^+,\pm3/2\rangle$, respectively; (c) and (d) the same for the sublevel $|5/2^+,\pm3/2\rangle$; (e) transitions that can accompany the process shown in the diagram (d) under condition $\Delta\omega_L>\Delta{E}_{gr_2}-\Delta{E}_{is}$.}
\label{fig:LT}
\end{figure}

The ground state and the low energy isomeric state in $^{229}$Th have quadrupole moments $Q_{gr} = 3.15$ $e$b \cite{Bemis-88} and $Q_{is} \simeq 1.8$ $e$b \cite{Tkalya-11}.  The interaction of the quadrupole moments with the electric field gradient leads to a splitting of the nuclear states to sublevels doubly degenerated in the sign of the magnetic quantum number $m$ (see in Fig.~\ref{fig:LT}).

For numerical estimates we consider two crystals where $^{229}$Th$^{4+}$ ions substitute one of the chemical elements with lower valence, namely, Ca$^{2+}$. These are crystals of LiCaAlF$_6$ \cite{Rellergert-10} and CaF$_2$ \cite{Kazakov-12} doped by $^{229}$Th. EFG at $^{229}$Th in $^{229}$Th:LiCaAlF$_6$ and $^{229}$Th:CaF$_2$ can be very large \cite{Tkalya-11} because of the contributions from two compensating ions F$^-$ located near the $^{229}$Th$^{4+}$ ion \cite{Jackson-09}. It can be estimated from the ratio
$$
\varphi_{zz}^{eff} = 2(1-\gamma_{\infty})\varphi_{zz},
$$
where $\varphi_{zz}$ is EFG from a single ion of F$^-$, and $\gamma_{\infty}$ is the so called Sternheimer quadrupole antishielding factor, whose estimate is $\gamma_{\infty}$ = -177.5 \cite{Feiock-69}.

%
%

As a result of the proximity of compensating ions to the thorium ion and the large antishielding factor, the effective electric field gradient at the $^{229}$Th nucleus can reach $\varphi_{zz}^{eff} \simeq-2\times10^{19}$ V cm$^{-2}$ (or $\simeq-2\times10^{20}$ eV cm$^{-2}$). Consequently, the interaction energy,
$$
E^{int}_{gr(is)} = e Q_{gr(is)} \varphi_{zz}^{eff} ,
$$
will be $6\times10^{-5}$ eV for the ground state and $3.3\times10^{-5}$ eV for the isomeric state. From the relation
$$
E^{(m)}_{gr(is)} = E^{int}_{gr(is)}\frac{3m^2-J_{gr(is)}(J_{gr(is)}+1)}{4J_{gr(is)}(2J_{gr(is)}-1)}
$$
it is easy to calculate the value of the splitting (see in Fig.~\ref{fig:LT}~(a)): $\Delta{E}_{gr_1}=0.9\times10^{-5}$ eV, $\Delta{E}_{gr_2}=1.8\times10^{-5}$ eV, and $\Delta{E}_{is}=1.6\times10^{-5}$ eV.

The estimates for the energy splitting of nuclear states are preliminary because there are no direct experimental data on the value of the quadrupole moment of the isomeric state $Q_{is}$ and electric field gradient at the $^{229}$Th nucleus in the crystals. However, they are obtained in the framework of generally accepted models and their qualitative values can be very useful for preparation of future experiments.

\par
\section{Depopulation of sublevels by narrowband laser}
\label{sec:Depopulation}
\par

All $M1$ transitions allowed by the selection rules in the $^{229}$Th ground-state doublet are shown in Fig.~\ref{fig:LT}. These transitions lead to the depopulation of the ground state sublevels via sublevels of the isomeric state. As an example below we consider in detail two of the presented depopulation schemes resulting in inverse population. All the schemes in Fig.~\ref{fig:LT} are similar, they are described by typical equations, but have different efficiency.

We start by considering depopulation of the middle sublevel $|5/2^+,\pm3/2\rangle$ according to the scheme in Fig.~\ref{fig:LT}~(c) when laser beam with the width $\Delta\omega_L\ll\Delta{E}_{gr_1}$ is tuned to the nuclear transition $|5/2^+,\pm3/2\rangle\leftrightarrow|3/2^+,\pm3/2\rangle$. The sublevel $|3/2^+,\pm3/2\rangle$ has three decay channels: back to the state $|5/2^+,\pm3/2\rangle$, and to $|5/2^+,\pm5/2\rangle$ and $|5/2^+,\pm1/2\rangle$. Under laser radiation the density of nuclei $n^{gr}_{\pm3/2}$ in $|5/2^+,\pm3/2\rangle$ will gradually decrease, and the density of nuclei in the upper sublevel $n^{gr}_{\pm1/2}$ and the lower sublevel $n^{gr}_{\pm5/2}$ of the ground state will increase.

The equations describing this process can be written in the following form
\begin{equation}
\begin{array}{l}
dn^{gr}_{\pm1/2}/dt  =  d_{31} \Lambda{} n^{is}_{\pm3/2}\,,\\
dn^{gr}_{\pm3/2}/dt  = -u_{33}\sigma \varphi{} n^{gr}_{\pm3/2} + d_{33}(\Lambda + g\sigma\varphi) n^{is}_{\pm3/2}\,,\\
dn^{gr}_{\pm5/2}/dt  =  d_{35} \Lambda{} n^{is}_{\pm3/2}\,,\\
dn^{is}_{\pm3/2}/dt  =  u_{33} \sigma \varphi{} n^{gr}_{\pm3/2} - (\Lambda + d_{33} g\sigma\varphi) n^{is}_{\pm3/2}\,.
\end{array}
\label{eq:LT-c}
\end{equation}

The initial conditions,
\begin{equation}
\begin{array}{l}
n^{gr}_{\pm1/2}(0) = n^{gr}_{\pm3/2}(0) = n^{gr}_{5/2}(0) =1\,,\\
n^{is}_{\pm1/2}(0) = n^{is}_{\pm3/2}(0) =0
\end{array}
\label{eq:IC}
\end{equation}
correspond to a uniform population of the ground state with the {\it{unit}} density of $^{229}$Th in each of the sublevels and the unpopulated isomeric level. (One must remember that according to the initial conditions (\ref{eq:IC}) the total number of $^{229}$Th nuclei implanted in crystal, is normalized to 3. Below we will see that the normalization (\ref{eq:IC}) turns out to be more convenient for visual presentation of results of calculations than the condition $n^{gr}_{\pm1/2}(0) = n^{gr}_{\pm3/2}(0) = n^{gr}_{5/2}(0) =1/3$.)

Coefficients $u_{ij}$ in (\ref{eq:LT-c}) correspond to transitions to levels with higher energies and $d_{ij}$ to levels with lower energies.  They are given by the expression
$$
\frac{1}{2}\frac{2J_i+1}{2J_f+1}\sum_M \left(C^{J_f M_f}_{J_i M_i LM}\right)^2
$$
and take into account the properties of each of the partial transitions between the states $|5/2^+\rangle$ and $|3/2^+\rangle$. (Factor 1/2 is a consequence of the double degeneracy of the states with respect to the magnetic quantum number $m$. The initial and final state for the partial transitions shown in Fig.~\ref{fig:LT} are denoted by $i$ and $f$.) The numerical values of $u_{ij}$ and $d_{ij}$ are given in Table~\ref{tab:Coefficients}.

\begin{table}
\caption{\label{tab:Coefficients}Coefficients $u_{ij}$ and $d_{ij}$.}
\begin{indented}
\item[]\begin{tabular}{@{}llcc}
\br
    Initial state         &   Final state        & Coefficient & Numerical value\\
\mr
 $|5/2^+,\pm1/2\rangle$   &  $|3/2^+,\pm1/2\rangle$ & $u_{11}$ & 9/10 \\
 $|5/2^+,\pm1/2\rangle$   &  $|3/2^+,\pm3/2\rangle$ & $u_{13}$ & 1/10 \\
 $|5/2^+,\pm3/2\rangle$   &  $|3/2^+,\pm1/2\rangle$ & $u_{31}$ & 3/5  \\
 $|5/2^+,\pm3/2\rangle$   &  $|3/2^+,\pm3/2\rangle$ & $u_{33}$ & 2/5  \\
 $|5/2^+,\pm5/2\rangle$   &  $|3/2^+,\pm3/2\rangle$ & $u_{53}$ &  1   \\
 $|3/2^+,\pm1/2\rangle$   &  $|5/2^+,\pm1/2\rangle$ & $d_{11}$ & 3/5  \\
 $|3/2^+,\pm1/2\rangle$   &  $|5/2^+,\pm3/2\rangle$ & $d_{13}$ & 2/5  \\
 $|3/2^+,\pm3/2\rangle$   &  $|5/2^+,\pm1/2\rangle$ & $d_{31}$ & 1/15 \\
 $|3/2^+,\pm3/2\rangle$   &  $|5/2^+,\pm3/2\rangle$ & $d_{33}$ & 4/15 \\
 $|3/2^+,\pm3/2\rangle$   &  $|5/2^+,\pm5/2\rangle$ & $d_{35}$ & 2/3 \\
\br
\end{tabular}
\end{indented}
\end{table}

The cross section of nuclear photoexcitation is calculated \cite{Tkalya-11} from the standard relation
$$
\sigma = \frac{\lambda_{is}^2}{2\pi} \frac{\Gamma_{rad}/g}{\Delta\omega_L},
$$
where $\lambda_{is}=2\pi/E_{is}$ is the wavelength, and $\Gamma_{rad}$ is the radiative width of the isomeric transition $|3/2^+\rangle\rightarrow|5/2^+\rangle$.

In our case $\Gamma_{rad}=\Lambda$. A preliminary value of $\Gamma_ {rad} \simeq3\times10^{-19}$ eV for the decay of the isomer $^{229}$Th$^m$($3/2^+,7.6$ eV in a crystal with a wide band gap and the refractive index close to 1 was obtained  in \cite{Tkalya-11}. If the width of the laser line satisfies the condition $\Delta\omega_L\ll\Delta{}E_{gr_{2}}-\Delta{}E_{is}$ (for this case, we used the value of $\Delta\omega_L\sim10^{-7}$ eV for numerical estimations) than the laser radiation excites selectively the nucleus only from the required sublevel without affecting the other sublevels. If the width satisfies the condition $\Delta\omega_L\geq \Delta{}E_{gr_{2}}-\Delta{}E_{is}$ (i.e. $\Delta\omega_L\sim10^{-6}$ eV for example) than we have simultaneous excitation of the nuclei $^{229}$Th from the two lower sublevels through the transitions $|5/2^+,\pm3/2\rangle\rightarrow|3/2^+,\pm1/2\rangle$ and $|5/2^+,\pm5/2\rangle\rightarrow|3/2^+,\pm3/2\rangle$ (see excitation schemes (d) and (e) in Fig.~\ref{fig:LT}).

\begin{figure}[]
\begin{center}
\epsfig{file=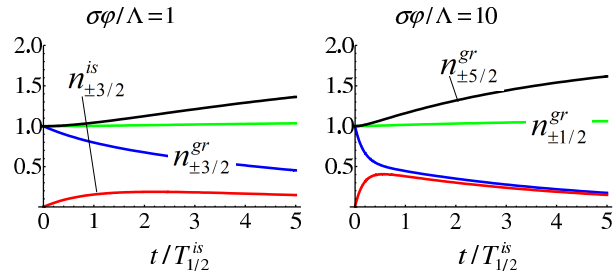,width=8.0cm}
\end{center}
\caption{Time evolution of the populations of the state $|5/2^+,\pm3/2\rangle$ in the irradiation scheme of Fig.~\ref{fig:LT}~(c). Plots in the left and right panels are calculated for $\sigma\varphi/\Lambda=1$ and 10, respectively. (Since the relative positions of the plots are the same the labels are not duplicated.)}
\label{fig:Depopulation-c}
\end{figure}

Time evolution of the density of nuclei in these sublevels [shown in Fig.~\ref{fig:LT}~(c) and described by (\ref{eq:LT-c})] is shown in Fig.~\ref{fig:Depopulation-c}. Left and right panels correspond to $\sigma\varphi/\Lambda=1$ and $\sigma\varphi/\Lambda=10$, respectively. For the chosen $\Gamma_{rad}$ the decay constant is $\Lambda\simeq4.5\times10^{-4}$ s$^{-1}$. Furthermore, for $\Delta\omega_L\sim10^{-7}$ eV the excitation cross section is $\sigma\simeq10^{-22}$ cm$^{-2}$. Therefore, the parameters $\sigma\varphi/\Lambda=1$ and $\sigma\varphi/\Lambda=10$ correspond to realistic values $\varphi \simeq10^{19}$ cm$^{-2}$s$^{-1}$ and $\varphi \simeq10^{20}$ cm$^{-2}$s$^{-1}$. Laser with the average power of 10 mW provides these photon flux density through the target surface area of 10$^{-3}$ cm$^2$ and 10$^{-4}$ cm$^2$ respectively. From Fig.~\ref{fig:Depopulation-c} it is clear that the process of depopulation of the state $|5/2^+,\pm3/2\rangle$ via the sublevel $|3/2^+,\pm3/2\rangle$ occurs relatively quickly, especially for $\sigma\varphi/\Lambda=10$.

%
%

\par
\section{Creation of inverse population}
\label{sec:IP}
\par

Now we turn to the second stage of the process, which goal is to obtain the inverse population of nuclear levels.
Since according to Fig.~\ref{fig:Depopulation-c}, the majority of nuclei was transformed from $|5/2^+,\pm3/2\rangle$ to $|5/2^+,\pm5/2\rangle$, it is logical to consider subsequent excitations to the isomeric level from the state $|5/2^+,\pm5/2\rangle$.

Therefore, we further adopt the scheme shown in Fig.~\ref{fig:LT}~(e). As initial conditions for this scheme we use the densities $n^{gr}_{\pm1/2}$, $n^{gr}_{\pm3/2}$, $n^{gr}_{\pm5/2}$ and $n^{is}_{\pm3/2}$ obtained in the output of the calculation of (\ref{eq:LT-c}) at time $t=10T_{1/2}^{is}$ with the initial conditions (\ref{eq:IC}). The system of equations for the scheme shown in Fig.~\ref{fig:LT}~(e) is written as
\begin{equation}
\begin{array}{l}
dn^{gr}_{\pm1/2}/dt  =  d_{31}\Lambda{} n^{is}_{\pm3/2}\,,\\
dn^{gr}_{\pm3/2}/dt  =  d_{33}\Lambda{} n^{is}_{\pm3/2}\,,\\
dn^{gr}_{\pm5/2}/dt  =  -u_{53} \sigma \varphi{} n^{gr}_{\pm5/2} + d_{35} (\Lambda + g\sigma\varphi) n^{is}_{\pm3/2}\,,\\
dn^{is}_{\pm3/2}/dt  =  u_{53} \sigma \varphi{} n^{gr}_{\pm5/2} - (\Lambda + d_{35} g\sigma\varphi) n^{is}_{\pm3/2}\,.
\end{array}
\label{eq:LT-e}
\end{equation}

\begin{figure}[h]
\begin{center}
\epsfig{file=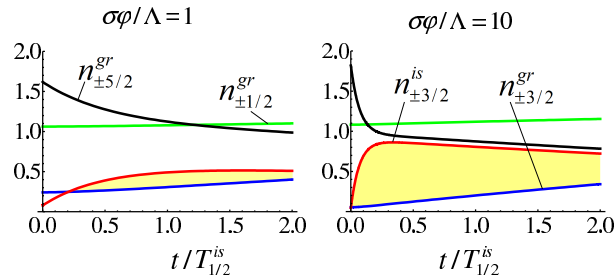,width=8.0cm}
\end{center}
\caption{Inverse population between the sublevels $|3/2^+,\pm3/2\rangle$ and $|5/2^+,\pm3/2\rangle$, obtained as a result of the consecutive processes (c) and (e) shown in Fig.~\ref{fig:LT}.}
\label{fig:IP}
\end{figure}

Numerical results for (\ref{eq:LT-e}) are presented in Fig.~\ref{fig:IP}. If at two stages $\sigma\varphi/\Lambda=1$, we see that the inverse population $n^{is}_{\pm3/2}-n^{gr}_{\pm3/2}$ reaches the value of $\simeq0.1$ at time $T_{1/2}^{is}$. In the case $\sigma\varphi/\Lambda=10$ we have $n^{is}_{\pm3/2}-n^{gr}_{\pm3/2}\simeq0.8$ already at $t\simeq0.2T_{1/2}^{is}$.

%
%

These results lead to the following estimations for real systems. In the work \cite{Rellergert-10} it was reported that the density of $^{229}$Th nuclei implanted into the LiCaAlF$_6$ crystal can achieve $n^{gr}_0\simeq10^{18}$ cm$^{-3}$. Thus, the initial density of nuclei in our model (adopted as 1 in our calculations)
should be $n^{gr}_0/3\simeq3\times10^{17}$ cm$^{-3}$. Correspondingly, after two stages of irradiation with the parameter $\sigma\varphi/\Lambda=10$, the inverse population $n^{is}_{\pm3/2}-n^{gr}_{\pm3/2}$ in such a crystal could be $2.5\times10^{17}$ cm$^{-3}$. As has been shown in \cite{Tkalya-11}, this is sufficient for effective amplification of electromagnetic radiation by nuclear matter, that is, for the creation of a VUV band laser on nuclear transition.

\par
\section{Simultaneous depopulation of two sublevels}
\label{sec:WL}
\par

Let us consider the situation when the energy splitting $\Delta{}E_{is}-\Delta{}E_{gr_2}$ is smaller than the width of laser line, $\Delta\omega_L$. In that case the transition processes shown in Fig.~\ref{fig:LT}~(d) and (e) will occur simultaneously. The system of equations corresponding to this case (labeled as Fig.~\ref{fig:LT}~(d)+(e)) has the form
\begin{equation}
\begin{array}{l}
dn^{gr}_{\pm1/2}/dt  =  \Lambda(d_{11} n^{is}_{\pm1/2} + d_{31} n^{is}_{\pm3/2})\,,\\
dn^{gr}_{\pm3/2}/dt  =  -u_{31} \sigma \varphi{} n^{gr}_{\pm3/2} + d_{13} (\Lambda + g\sigma\varphi) n^{is}_{\pm1/2} +\\
\qquad{} d_{33}\Lambda{} n^{is}_{\pm3/2}\,,\\
dn^{gr}_{\pm5/2}/dt  =  -u_{53} \sigma \varphi{} n^{gr}_{\pm5/2} + d_{35} (\Lambda + g\sigma\varphi)n^{is}_{\pm3/2}\,,\\
dn^{is}_{\pm1/2}/dt  =  u_{31} \sigma \varphi{} n^{gr}_{\pm3/2} - (\Lambda + d_{13} g\sigma\varphi) n^{is}_{\pm1/2}\,,\\
dn^{is}_{\pm3/2}/dt  =  u_{53} \sigma \varphi{} n^{gr}_{\pm5/2} - (\Lambda + d_{35} g\sigma\varphi) n^{is}_{\pm3/2}\,.
\end{array}
\nonumber
\label{eq:LT-f}
\end{equation}

\begin{figure}[h]
\begin{center}
\epsfig{file=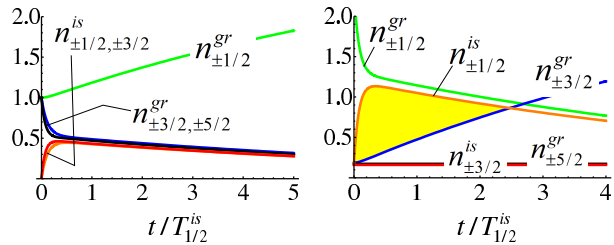,width=8.0cm}
\end{center}
\caption{Simultaneous depopulation of the states $|5/2^+,\pm3/2\rangle$ and $|5/2^+,\pm5/2\rangle$ through transitions shown in Fig.~\ref{fig:LT}~(d)+(e) in the case $\sigma\varphi/\Lambda=10$ (left panel) and creation of inverse population according to the scheme shown in Fig.~\ref{fig:LT}~(b) (right panel).}
\label{fig:D-IP}
\end{figure}

Numerical solution for that case is given in Fig.~\ref{fig:D-IP} (plots in the left panel). It is clearly seen that the population of the level $|5/2^+,\pm1/2\rangle$ grows quickly. For a time of (8--10)$T^{is}_{1/2}$ it doubles. At the same time populations of all other states are reduced. Now one can irradiate the sample according to the scheme in Fig.~\ref{fig:LT}~(b) and reach the inverse population $n^{is}_{\pm1/2}-n^{gr}_{\pm3/2}\simeq1$  already at $t\simeq0.2T_{1/2}^{is}$ (see plots in the right panel in Fig.~\ref{fig:D-IP}). A slightly less effect can be achieved by irradiation of the sample according to the scheme in Fig.~\ref{fig:LT}~(a). Calculations show that in this case $n^{is}_{\pm3/2}-n^{gr}_{\pm3/2}\simeq0.4$ at $t\simeq{}T_{1/2}^{is}$.

%
%

\par
\section{Excitation of nucleus by broadband laser with anti-Stokes scattering}
\label{sec:WL}
\par

In this section we consider the excitation of the nuclear isomer $^{229}$Th$^m$($3/2^+$, 7.6 eV) by broadband laser ($\Delta\omega_L>\Delta{}E_{is},\Delta{}E_{gr_{1,2}}$) taking into account the anti-Stokes scattering. This process is described by the following equations for nuclear populations
\begin{equation}
\begin{array}{l}
dn^{gr}_{\pm1/2}/dt  = -\sigma\varphi{}n^{gr}_{\pm1/2} + (\Lambda+g\sigma\varphi)\times \\
\qquad (d_{11}n^{is}_{\pm1/2}+d_{31}n^{is}_{\pm3/2})\,,\\
dn^{gr}_{\pm3/2}/dt  = -\sigma\varphi{}n^{gr}_{\pm3/2}+ (\Lambda + g\sigma\varphi)\times \\
\qquad (d_{13}n^{is}_{\pm1/2} + d_{33}n^{is}_{\pm3/2})\,,\\
dn^{gr}_{\pm5/2}/dt  = -\sigma\varphi{}n^{gr}_{\pm5/2} + d_{35}(\Lambda+g\sigma\varphi)n^{is}_{\pm3/2}\,,\\
dn^{is}_{\pm1/2}/dt  = \sigma\varphi(u_{11}n^{gr}_{\pm1/2} + u_{31}n^{gr}_{\pm3/2})- \\
\qquad (\Lambda+g\sigma\varphi)n^{is}_{\pm1/2}\,,\\
dn^{is}_{\pm3/2}/dt  = \sigma\varphi(u_{13}n^{gr}_{\pm1/2} + u_{33}n^{gr}_{\pm3/2} +  \\
\qquad{} u_{53}n^{gr}_{\pm5/2}) - (\Lambda+g\sigma\varphi)n^{is}_{\pm3/2}\,.
\end{array}
\label{eq:WB}
\end{equation}
Here we have taken into account that $u_{11}+u_{13}=u_{31}+u_{33}=d_{11}+d_{13}=d_{31}+d_{33}+d_{35}=u_{53}=1$.

\begin{figure}[h]
\begin{center}
\epsfig{file=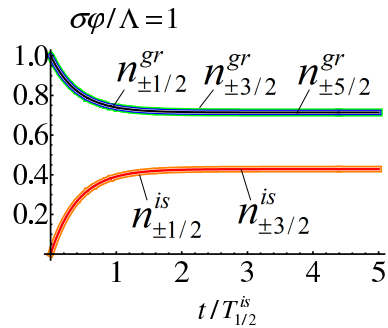,width=5.0cm}
\end{center}
\caption{Nuclear densities of sublevels after irradiation by broadband laser when anti-Stokes scattering is taken into account.}
\label{fig:WB}
\end{figure}

Numerical solutions of (\ref{eq:WB}) with the initial conditions (\ref{eq:IC}) are presented in Fig.~\ref{fig:WB}. Note, that the time dependencies of nuclear densities in various sublevels of the ground state virtually coincide. The same applies to the population of the sublevels in the isomeric state. The relation $n^{gr}_{\pm1/2}+n^{gr}_{\pm3/2}+n^{gr}_{\pm5/2}+n^{is}_{\pm1/2}+n^{is}_{\pm3/2}=3$ is satisfied at any given time. (We recall that the occupation number here is 3 as a consequence of the initial conditions (\ref{eq:IC}).) The resulting dependencies of Fig.~\ref{fig:WB} confirm the conclusion of \cite{Tkalya-11} that the inverse population of $^{229}$Th nucleus can not be obtained by means of broadband laser radiation.

%
%

\section{Conclusion}
In conclusion, a sufficiently effective and relatively fast method for creating of inverse population of $^{229}$Th nuclei by means of radiation of narrowband laser is proposed. The inverse population obtained between the quadrupole sublevels of the low energy isomeric level 3/2$^+$(7.6 eV) and the ground state 5/2$^+$(0.0) of the $^{229}$Th nucleus is reached for a time of the order of 10$T_{1/2}^{is}$. The inverse population is sufficient for enabling the amplification of electromagnetic radiation with the photon energy of 7.6 eV by the medium containing the isomeric nuclei $^{229}$Th$^m$.

\ack
The authors would like to thank Dr. A.~V.~Nikolaev for useful discussions.

\end{document}